\documentclass{llncs}

\usepackage{epsfig}
\usepackage{graphicx}
\usepackage[table,xcdraw]{xcolor}
\usepackage{listings}

\begin{document}

\title{Spark Parameter Tuning via Trial-and-Error}
\author{Panagiotis Petridis\inst{1} \and Anastasios Gounaris\inst{1} \and Jordi Torres\inst{2}}
\institute{Dept. of Informatics,
Aristotle University of Thessaloniki, Greece\\
\email{\{ppetridis,gounaria\}@csd.auth.gr}
\and
Computer Architecture Dept., Technical University of Catalonia, Spain\\ \email{torres@ac.upc.edu}}

\maketitle

\begin{abstract}
Spark has been established as an attractive platform for big data analysis, since it manages to hide most of the complexities related to parallelism, fault tolerance and cluster setting from developers. However, this comes at the expense of having over 150 configurable parameters, the impact of which cannot be exhaustively examined due to the exponential amount of their combinations. The default values allow developers to quickly deploy their applications but leave the question as to whether performance can be improved open.
In this work, we investigate the impact of the most important of the tunable Spark parameters on the application performance and guide developers on how to proceed to changes to the default values. We conduct a series of experiments with known benchmarks on the MareNostrum petascale supercomputer  to test the performance sensitivity. More importantly, we offer a trial-and-error methodology for tuning parameters in arbitrary applications based on evidence from a very small number of experimental runs. We test our methodology in three case studies, where we manage to achieve speedups of more than 10 times.
\end{abstract}

\section{Introduction}

Spark \cite{zaharia2012,spark} has emerged as one of the most widely used frameworks for massively parallel data analytics.  In summary, it improves upon Hadoop MapReduce in terms of flexibility in the programming model and performance \cite{ShiQMJWRO15}, especially for iterative applications. It can accommodate both batch and streaming applications, while providing interfaces to other established big data technologies, especially regarding storage, such as HDFS and NoSQL databases. Finally, it includes components for SQL-like processing, graph processing, machine learning and data mining. However, its key feature  is that it manages to hide the complexities related to parallelism, fault-tolerance and cluster setting from end users and application developers. This feature renders Spark practical for use in real-life data science and big data processing applications.

To support all the features mentioned above, Spark execution engine has been evolved to an efficient albeit complex system with more than 150 configurable parameters. The default values are usually sufficient for a Spark program to run, e.g., not to run out of memory without having the option to spill data on the disk and thus crash. But this gives rise to the following research question: \emph{``Can the default configuration be improved?"}

The aim of this work is to answer the above question in an efficient manner. Clearly, it is practically impossible to check all the different combinations of parameter values for all tunable parameters. Therefore, tuning arbitrary Spark applications by inexpensively navigating through the vast search space of all possible configurations in a principled manner is a challenging task.

Very few research endeavors focus on issues related to understanding the performance of Spark applications and the role of tunable parameters \cite{ousterhout2015making,awan2015data,tous2015spark}. For the latter, Spark's official configuration\footnote{\url{http://spark.apache.org/docs/latest/configuration.html}} and tuning\footnote{\url{http://spark.apache.org/docs/latest/tuning.html}} guides and tutorial book \cite{Spark-book} provide a valuable
asset in understanding the role of every single parameter.

Nevertheless, understanding the role of a parameter does not necessarily mean that the impact of each parameter on the performance of arbitrary applications is understood as well. Moreover, such an understanding does not imply that tuning is straightforward. An added complexity stems from the fact that most parameters are correlated  and the impact of  parameters may vary from application to application and it will also vary from cluster to cluster.

In this work, we experiment with the MareNostrum petascale supercomputer at the Barcelona Supercomputing Center in Spain. After configuring the cluster in an application-independent way according to the results in \cite{tous2015spark}, we examine the impact of configurable parameters on a range of applications and derive a simple trial-and-error tuning methodology that can be applied to each Spark application separately. We test our methodology using three case studies with particularly encouraging results.

%
%

The summary of our contributions is as follows. (i) We provide an overview of the known results to date on configuring Spark applications.
(ii) We identify the most important parameters in terms of their potential impact on performance and we assess this impact in real runs on the Marenostrum petascale supercomputer. The number of these parameters is 12. (iii) Based on our results, we propose a novel tuning methodology to be applied on an individual application basis (summarized in Fig. \ref{fig:diagram}). The methodology treats applications as black boxes, follows an efficient trial-and-error approach that involves a low number of experimental runs for just 10 different configurations at most, and takes into account the correlation between different parameters. (iv) We evaluate our methodology in practice using three case studies and we show that we can achieve significant speedups (up to more than 10 times).

The remainder of this work is structured as follows. The next section provides an overview of the known results to date with regards to Spark tuning. In Sec. \ref{sec:params}, we explain the chosen parameters. The experiments that assess the sensitivity of performance to the parameters under investigation are presented in Sec. \ref{sec:exps}. Sec. \ref{sec:methodology} deals with our proposed tuning methodology and the evaluation of its effectiveness. We discuss conclude in Sec. \ref{sec:concl}.

\section{Overview of Existing Results for Spark Configuration}
\label{sec:overview}


\begin{table}[tb!]
\centering
\scriptsize
\begin{tabular}{|p{0.15\textwidth}|p{0.8\textwidth}|}
\hline
Category & Description \\ \hline
\emph{Application Properties} & The parameters in this group concern basic application properties, such as the application name, the CPU and the memory resources  that will be allocated to the executor and driver components, and  so on. \\ \hline

\emph{Runtime Environment} & The parameters belonging to this group refer to environment settings, such as classpaths, java options and logging.
\\ \hline
\emph{Shuffle Behavior} & These parameters have to do with the shuffling mechanism of Spark, and  they involve buffer settings, sizes, shuffling methods, and so on.
\\ \hline
\emph{Spark UI} & The UI parameters are mostly related to UI event-logging.
\\ \hline
\emph{Compression and Serialization} &  The parameters of this group target compression and serialization issues. They mostly have to do with whether compression will take place or not, what compression codec will be used, the codec's parameters, the serializer to be used and its buffer options.
\\ \hline
\emph{Memory Management} & These parameters fix the fractions of memory (heap size) allocated to storage, shuffling and execution through at a fine-granularity level.
\\ \hline
\emph{Execution Behavior} & The parameters of this group refer to a wide range of execution details including the number of execution cores and data parallelism.
\\ \hline
\emph{Networking} & This group contains parameters that provide options for network issues.  Most parameters refer to timeout options, heartbeat pauses, ports and network retries. \\ \hline
\emph{Scheduling} & Most parameters in the group cover scheduling options. Some noteworthy parameters have to do with scheduling mode, and whether to use the speculation optimization and the maximum number of CPU cores that will be used.
\\ \hline
\emph{Dynamic Allocation }& By using dynamic allocation, Spark can increase or decrease the number of executors based on its workload. This is only available in Yarn (which we do not use in this work).
\\ \hline
\emph{Security} & These parameters deal with authentication issues.
\\ \hline
\emph{Encryption} & These properties refer to encryption algorithms, passwords and keys that may be employed.
\\ \hline
\emph{Spark Streaming and SparkR} & These parameters are specific to the  Spark Streaming and SparkR higher-level components.
\\ \hline
\end{tabular}
\caption{Summary of parameter categories}
\label{tab:categories}
\vspace{-0.5cm}
\end{table}

Table \ref{tab:categories} provides a categorization of Spark parameters. In this work, we target parameters belonging to the \emph{Shuffle Behavior, Compression and Serialization}, and \emph{Memory Management} categories. Note that there are several other parameters belonging to categories such as \emph{Application Properties, Execution Behavior} and \emph{Networking} that may affect the performance, but these parameters are typically set at the cluster level, i.e., they are common to all applications running on the same cluster of machines, e.g., as shown in \cite{tous2015spark}.

Next, we summarize the known results to date with regards to Spark configuration. These results come from three types of sources: (i) academic works that aimed at Spark configuration and profiling; (ii) official Spark documentation and guides that build on top of this documentation; and (iii) academic works that include evaluation of Spark applications on real platforms and, as a by-product, provide information about the configuration that yielded the highest performance. We also briefly discuss results on Spark profiling, because they directly relate to Spark configuration. The most relevant work to ours is the study of Spark performance on the MareNostrum supercomputer in Barcelona in \cite{tous2015spark}, which is presented separately.

{\bf Optimization of Spark on MareNostrum.} The work in \cite{tous2015spark} shed lights onto the impact of configurations related to parallelism.
In MareNostrum, cluster management is performed according to the standalone mode, i.e., YARN and MESOS are not used.
The main results of \cite{tous2015spark}, which are reused in our work, are summarized as follows. First,
the number of cores allocated to each Spark executor, which is responsible for running local Spark tasks in a massively parallel setting, has a big impact on performance, but should be configured in an application-independent manner. In other words, all applications sharing the same cluster can share the same corresponding configuration.  Second,
the level of parallelism plays a significant role. In cpu-intensive applications, such as k-means, allocating a single data partition per participating core yields the highest performance. In applications with a higher proportion of data-shuffling, such as sort-by-key, increasing the number of data partitions per core is recommended.  A range of additional aspects have been investigated in \cite{tous2015spark}, but their significance was shown to be small. These aspects included forcing the cores of a single executor to reside on the same physical machines, and using Ethernet instead of Infiniband network. Finally, it is discussed that the type of the underlying file system affects the performance, however, this is a property of the infrastructure rather than a tunable parameter.

{\bf Guides from Spark documentation.} Spark official documentation presents a summary of  tuning guidelines can be summarized as follows. (i)
The type of the serializer is an important configuration parameter. The default option uses Java's framework, but if \texttt{Kryo} library is applicable, it may reduce running times significantly.
(ii) The memory allocated to main computations and shuffling and the memory allocated to caching. It is stated that \emph{``although there are two relevant configurations, the typical user should not need to adjust them as the default values are applicable to most workloads"}. Memory-related configurations are also related to the performance overhead of garbage collection (GC).
(iii)
The level of parallelism, i.e., the number of tasks in which each RDD is split needs to be set in a way that the the cluster resources are fully utilized.
(iv) Data locality, i.e., enforcing the processing to take place where data resides, is important for distributed applications. In general, Spark scheduling respects data locality, but if a waiting time threshold is exceeded, Spark tasks are scheduled remotely. These thresholds are configurable.
Apart from the tuning guidelines above, certain other tips are provided by the Spark documentation, such as preferring arrays to hashmaps, using broadcasted variables, and mitigating the impact of GC through caching objects in serialized form. Also, similar guidelines are discussed more briefly  in \cite{Spark-book}.

In our work, we explicitly consider the impact of serialization and memory allocation. Tuning the  parallelism degree is out of our scope, but we follow the guidelines in \cite{tous2015spark}, so that resources are always occupied and data partitioning is set in a way that is efficient for the MareNostrum supercomputer. Also, we do not deal with GC and locality-related thresholds, because they have not seemed to play a big role in our experiments, where memory was  not a scarce resource.

Based on these guidelines, Alpine Data has published online a so-called \emph{cheat-sheet}, which is a tuning guide for system administrators\footnote{available from \url{http://techsuppdiva.github.io/spark1.6.html}}. The guide has the form of a block diagram and is tailored to the YARN cluster manager. Compared to our tuning model, it is more complex, and contains checks from logs and tips to modify the application code and setting of configuration parameters per application. By contrast, we focus only on the latter aspect considering each application as a black box requiring significantly fewer comparisons and considering more tuning parameters.

{\bf Other sources.}
In \cite{awan2015data}, the impact of the input data volume on Spark's applications is investigated.  The key parameters identified were related to memory and compression, although their exact impact is not analyzed. By contrast, we examine a superset of these parameters. An interesting result of this study is that GC time does not scale linearly with the data size, which leads to performance degradation for large datasets. In addition,
it is shown that if the input data increases, the idle cpu time may increase as well. The work in \cite{wang2014} is similar to \cite{tous2015spark} in that it discusses the deployment of Spark on a high-performance computing (HPC) system. In the described its evaluation, it identifies also four key Spark parameters along with the application-dependent level of parallelism. All these parameters are included in our discussion.

Finally, the work in \cite{davidson2013optimizing} focuses on Spark's shuffling optimization using two approaches. The first one is through columnar compression, which however does not yield any significant performance improvement. The second approach employs file consolidation  during shuffling. We consider file consolidation as well.

{\bf Profiling Spark applications and other related work.}
A thorough investigation  of bottlenecks in Spark is presented in \cite{ousterhout2015making}. The findings are also very interesting, since it is claimed that many applications are CPU-bound and memory plays a key role. In our experiments, we perform memory fine-tuning, since we also identify memory management as one of the most important parameters.

In \cite{chung2004using}, the authors present their tuning methodology, termed as Active Harmony, which is an  automated runtime performance tuning system was developed.
Also, the work in \cite{holl2014new} deals with the issue of parameter optimization in workflows,  
but may involve far too many experimental runs, whereas we advocate a limited number of configuration runs independently from the application size.

\section{Parameters of Interest}
\label{sec:params}

Navigating through the vast search space is one the biggest challenges in parameter testing and tuning, due to the exponential increase of different configurations in the number of properties and their valid values. Based on evidence from (i)  the documentation and the earlier works presented  in Sec. \ref{sec:overview} and (ii) our own runs (for which we do not present results due to space constraints), we narrow our focus on 12 parameters, the configuration of which needs to be investigated according to each application instance separately. As already explained, the application-independent parameters that need to be specific for a specific data-center and those related to data parallelism are out of our scope. Finally, we do not consider parameters related to YARN or MESOS.

\begin{enumerate}
\item \texttt{spark.reducer.maxSizeInFlight:} If this value is increased, reducers would request bigger output chunks. This would increase the overall performance but may aggravate the memory requirements. So, in clusters that there is adequate memory and where the application is not very memory-demanding, increasing this parameter could yield better results. On the other hand, if a cluster does not have adequate memory available, reducing the parameter should yield performance improvements.

\item \texttt{spark.shuffle.compress:}  In general, compressing data before they are transferred over the network is a good idea, provided that the time it takes to compress the data and transfer them is less than the time it takes to transfer them uncompressed. But if transfer times are faster than the CPU processing times, the main bottleneck of the application is shifted to the CPU and the process is not stalled by the amount of data that are transferred over the network but from the time it takes for the system to compress the data. Clearly, the amount of data transmitted during shuffling is application-dependent, and thus this parameter must not be configured to a single value for all applications.

\item \texttt{spark.shuffle.file.buffer:}  The role of this parameter bears similarities to the \texttt{spark.shuffle.maxSizeInFlight} parameter, i.e., if a cluster has adequate memory, then this value could be increased in order to get higher performance. If not, there might be performance degradation, since too much memory would allocated to buffers.

\item \texttt{spark.shuffle.manager:}    The available implementations are three: \emph{sort, hash} , and \emph{tungsten-sort}. \emph{Hash}  creates too many open files for certain inputs and aggravates memory limitations. However, combining the enabling of the \texttt{spark.shuffle.consolidateFiles} parameter with the \textit{Hash} manager, may mitigate this problem. \textit{Tungsten-sort} is reported to yield the highest performance in general provided that certain requirements are met.\footnote{\url{https://issues.apache.org/jira/browse/SPARK-7081}} Overall, there is no clear winner among the shuffle manager options.

\item \texttt{spark.io.compression.codec:}   Three options are available, namely \emph{snappy, lz4,} and \emph{lzf}. Although there are many tests conducted by various authors (e.g., \cite{davidson2013optimizing}) for the generic case, the best performing codec is application-dependent.
\item \texttt{spark.shuffle.io.preferDirectBufs:}   In environments where off-heap memory is not tightly limited, this parameter may play a role in performance.

\item \texttt{spark.rdd.compress:}   The trade-offs with regards to this parameter are similar to those for  shuffle compress. However, in this case, the trade-off lies between CPU time and memory.

\item \texttt{spark.serializer:}   In Spark's documentation, it is stated  that \emph{KryoSerializer} is thought to perform much better than the default Java Serializer when speed is the main goal. However, the Java serializer is still the default choice, so this parameter needs to be considered.

\item \texttt{spark.shuffle.memoryFraction:}   If, during shuffling, spills are often, then this value  should be increased from its default. Since this parameter is directly linked to the amount of memory that is going to be utilized, it may have a high performance impact. However, any increase is at the expense of the next parameter.

\item \texttt{spark.storage.memoryFraction:}  Since this parameter is directly linked to the amount of memory that is going to be utilized, it affects the performance.

\item \texttt{spark.shuffle.consolidateFiles:}  This parameter provides the option of consolidating intermediate files created during a shuffle, so that fewer files are created and performance is increased. It is stated however that, depending on the filesystem,  it may cause performance degradation.
\item \texttt{spark.shuffle.spill.compress:}   As for the previous compression options, a trade-off is involved.  If transferring the uncompressed data in an I/O operation is faster than compressing and transferring compressed data, then this option should be set to false. Provided that there is a high amount of spills, this parameter may have an impact on performance.
\end{enumerate}


\section{Sensitivity Analysis}
\label{sec:exps}

We employ three benchmark applications: (i) sort-by-key; (ii) shuffling and (iii) k-means.\footnote{The code of the applications can be found at \url{https://github.com/rtous/bsc.spark}} \textit{K-means} and \textit{Sort-by-key} are also part of the HiBench\footnote{https://github.com/intel-hadoop/HiBench} benchmark and where selected because they can be considered as representative of a variety of applications. The \emph{shuffling} application generates the data according to the terasort benchmark, but does not perform any sorting; it just shuffles all the data in order to stress the shuffling component of the system, given that shuffling in known to play a big role in the performance of Spark applications. To avoid the interference of the underlying file system, in all applications, the dataset was generated at the beginning of each run on the fly. The MareNostrum hardware specifications are described in \cite{tous2015spark}. 20 16-core machines are used, and the average allocated memory per core is 1.5GB. The version of Spark at the time of the experiments was 1.5.2.

For each of the selected parameters, we perform a separate set of tests, and in each test we examine a different value. Then, the performance is compared with the performance of the default configuration after modifying the serializer, as argued below. If there is a big difference between the results, then the parameter can be considered as having an impact on the overall performance.

The parameter values are selected as follows. If the parameter takes a binary value, for instance a parameter that specifies whether to use a feature or not, then the non-default value is tested. For parameters that have a variety of different values that are distinct, for instance the compression codec that will be used (\emph{snappy, lzf, lz4}), all the different values are tested. Finally, for parameters that take numeric values in a wide range, e.g, \texttt{spark.io.file.buffer}, the values close to the default are tested.
Each experiment was conducted five times (at least) and the median value is reported. 

\subsubsection{Sort-by-Key experiments.}
The setup of the \emph{sort-by-key} experiments is as follows. 1 billion key-value pairs are used, and each key and value have a length of 10 and 90 bytes, respectively. Also there are both 1000000 unique values for both keys and values. The degree of partitioned is set to 640, which is the optimal configuration according to the results in \cite{tous2015spark}.

\begin{figure}[tb!]
\centering
\includegraphics[width=0.99\columnwidth]{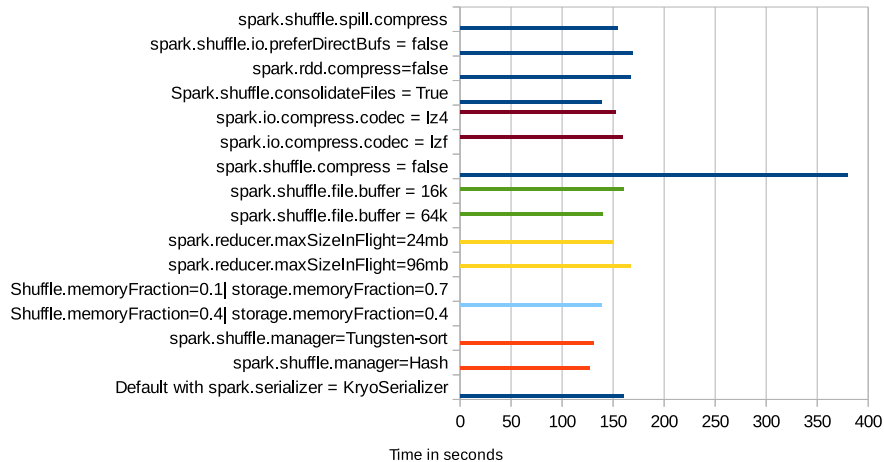}
\caption{Impact of all parameters for \emph{Sort-by-key}}
\label{srtbkall}
\vspace{-0.5cm}
\end{figure}

We first assess the impact of \texttt{spark.serializer}. \emph{KryoSerializer} performs significantly better than the default \emph{Java Serializer}, yielding approximately 25\% lower times. Since this gap is big, and in order to be able to extract insights regarding the performance of the rest of the parameters, the experiments that follow were conducted with the \emph{KryoSerializer}. The impact of all the remaining parameters is summarized in   Fig. \ref{srtbkall}. Different values of the same parameter are presented with the same color and as baseline, we use the \textit{KryoSerializer} performance, which is approximately 150 secs.

Starting the discussion from bottom to top, we see that both the non-default shuffle managers perform better than the default. \textit{Hash} performs at 127 seconds and \textit{Tungsten-sort} at 131 secs, nearly 30 secs faster than the default.
Regarding the memory fraction parameters, the values for \texttt{shuffle.memoryFraction} and \texttt{storage.memoryFraction} both set to 0.4 provide a less significant speedup (139 secs). One would expect these parameters to
have a bigger impact, given that \emph{sort-by-key} is a shuffling-intensive application. Interestingly, the second test for these parameters, with values of 0.1 and 0.7, respectively, led to application crash.
\texttt{spark.reducer.maxSizeInFlight} does not appear to be have a significant impact on performance. Increasing this parameter's value at 96mb yields the nearly same performance with the default (167 secs) but decreasing it to 24mb gives a small performance improvement (149 secs).
A similar observation can be drawn for \texttt{shuffle.file.buf-fer}; here, increasing the value yields slightly better results (140 secs).
The biggest impact on performance however can be seen in the \texttt{shuffle.compression} test runs. Disabling the compression degrades the performance by more than 100\%. It is worth noting that this may not always be the case, since the behavior of this parameter heavily relies on network and hardware details.
Regarding the compression codecs, there is not any noteworthy impact, since both \textit{lzf} and \textit{lz4} seem to be performing similarly to the default codec, \textit{snappy}. Also, the file consolidation implementation does not appear to provide any significant improvement either. This could be attributed to a variety of reasons, one being the fact that the \textit{sort-by-key} application does not generate a very high number of files during shuffling.
The last three parameters, \texttt{shuffle.spill.compress}, \texttt{shuffle.io.preferDirectBufs} and \texttt{rdd.compress} do not seem to significantly affect the performance too. For the former, this can be attributed to the fact that the spills conducted are few. For \texttt{Rdd.compress}, there is a small performance degradation as expected, since the RDD can fit into the main memory and CPU time is unnecessarily spent for  the compression.

\subsubsection{Shuffling experiments.}
In this set of experiments, the cluster setting was kept as in the previous one.
The raw dataset is of 400GB of data. Since the RDDs occupy more space, and the overall available memory in the executors is approximately 400GB for all RDDs and shuffling, accesses to the local storage are inevitable.



\begin{figure}[tb!]
\includegraphics[width=0.99\columnwidth]{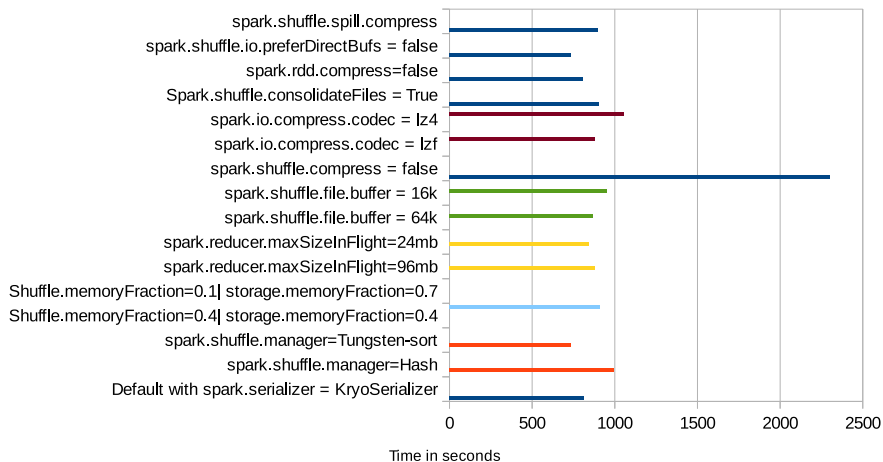}
\caption{Impact of all parameters for \emph{shuffling}}
\label{teraall}
\vspace{-0.5cm}
\end{figure}

Again, we first test the impact of \texttt{spark.serializer} and compare it with the default performance.
\textit{KryoSerializer} performs approx. 10\% better than the default Java serializer. In the experiments that follow, we use the performance of the \emph{KryoSerializer} as the baseline performance, which is 815 secs.
In Fig. \ref{teraall}, the results for the performance impact of the rest of the parameters are presented. As before, values of the same parameter are colored the same.

Contrary to the previous experiments, the \textit{Hash} shuffle manager performs worse resulting in performance degradation of 200 secs. This could probably be attributed to the fact that because the input much larger than the available memory for shuffling.  On the other hand, the \textit{Tungsten-sort} manager yields a speedup of 90secs. Similarly to \textit{sort-by-key}, increasing the memory available for shuffling at the expense of the \texttt{storage.memoryFraction} does not benefit the performance. Also, doing the opposite does not leave enough memory for shuffling and the application crashes.
Regarding the \texttt{reducer.maxSizeInFlight} parameter , there seems to be no significant impact on the application. Changing this parameter seems to affect  systems that have very small amounts of memory available. The same applies to \texttt{shuffle.file.buffer} parameter. The only difference however is that, when reducing the buffer size from 32KB to 15KB, the performance degrades by about 135 secs, which is more than 10\% from the initial execution. This might happen for the following reason. The buffers reduce the number of disk seeks and system calls made during the  creation of intermediate shuffle files. When the input is big enough and the buffers are relatively small, the number of system calls and disk seeks increases thus leading to performance degradation.
In addition, disabling the shuffle compression offers no improvement and greatly increases completion time.  Regarding the compression codec parameter, the \emph{lzf} codec does not seem to have any impact on the overall performance. However, the \emph{lz4} codec increases the application completion time by about 200 secs, i.e, incurring 25\% extra overhead.
Finally, the last three parameters configuration produce runtimes near the default value as was the case in the \textit{sort-by-key} experiments. As such, they considered as not having any significant impact on the performance. As stated before though, this may be specific to the infrastructure used, and in other environments, they may behave differently.

\subsubsection{K-means experiments.}



\begin{figure}[tb!]
\begin{tabular}{c}
\includegraphics[width=0.98\columnwidth]{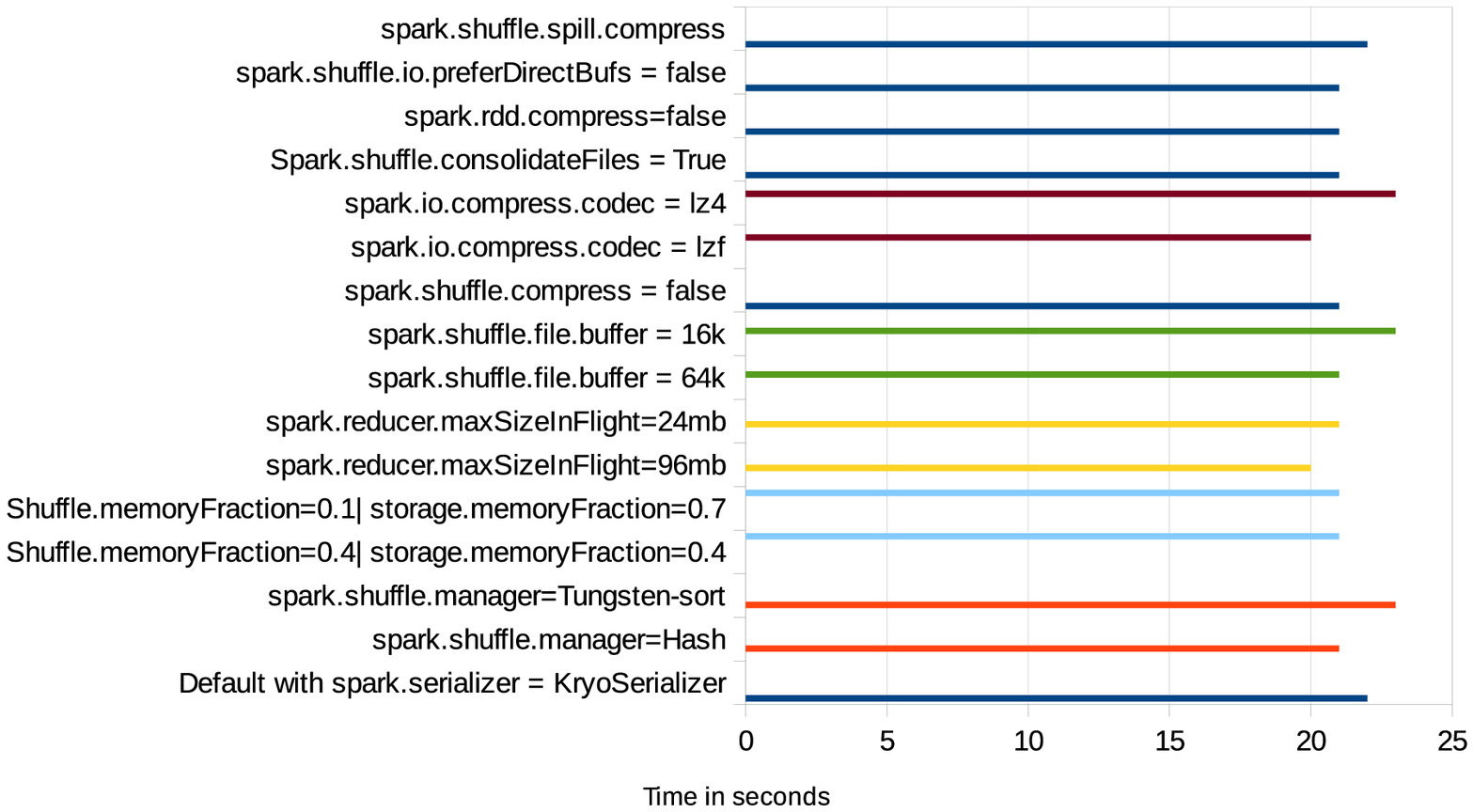} \\
\includegraphics[width=0.98\columnwidth]{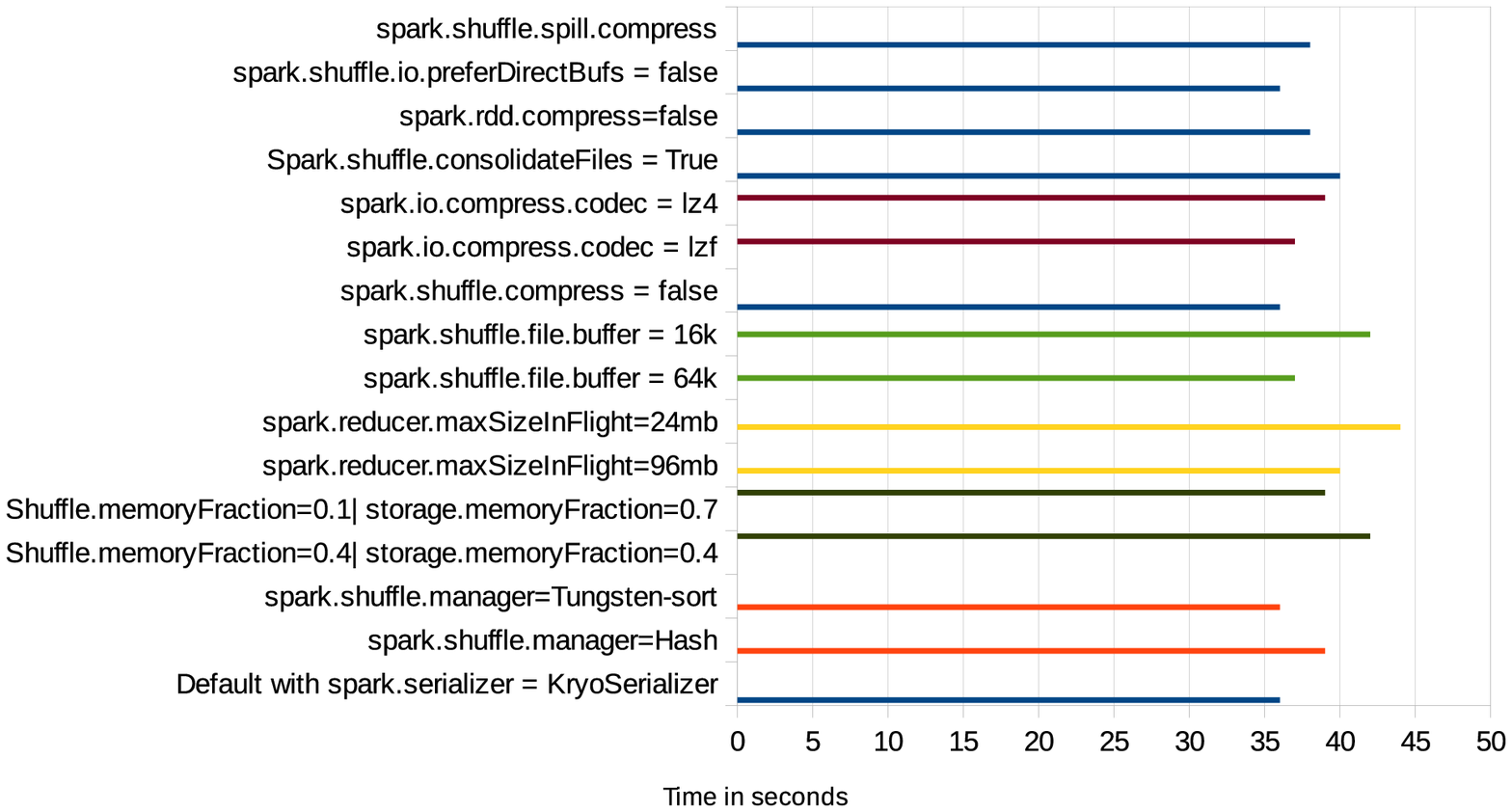}
\end{tabular}
\caption{Impact of all parameters for \emph{k-means} for 100M points(top) and 200M points (bottom)}
\label{kmall}
\vspace{-0.5cm}
\end{figure}

The final series of experiments tests the impact of the parameters on the performance of \emph{k-means}. We experimented with two data inputs with 100M and 200M 100-dimensional records, respectively. The number of centers was 10 and the amount of iterations was fixed to 10 to allow for fair time measurements.
In these cases, the impact of the \textit{KryoSerializer} (chosen as default) is very small.
The impact of  all the other parameters in Fig. \ref{kmall}.

An initial observation is that the impact of all parameters is less significant than in the previous two applications. From Fig. \ref{kmall}(top) we can see that, in absolute times, the differences are at most 2 secs, which is less than 10\%. This also applies, to a slightly less extent, to Fig. \ref{kmall}(bottom) as well, where the differences are up to 3 secs. However, although we conclude that the parameters selected do not affect the performance of the \textit{k-means} benchmark significantly, we can note the fact that the parameter \texttt{shuffle.compress} that dramatically degraded the performance in the previous experiments, has no overall impact on the test \emph{k-means} instances. This is actually expected, since it mostly affects the data shuffling part of the application, which plays a small, non-dominant role in \textit{k-means}. The same reasoning, i.e., data shuffling is not prominent in \emph{k-means}, explains the less impact of the other parameters, too.


\subsubsection{Overall Statistics.}
\par In Table \ref{avgimpact} the average impact of each parameter can be seen for each benchmark. All percentages refer to the mean deviation from the default runtime, regardless of whether the deviation is for the better or worse performance. The default is enabling the \emph{KryoSerializer}, apart from when testing the impact of this serializer itself. The lowest quartile of the parameters in terms of the magnitude of their incurred average impact are disregarded in the following tuning methodology, so that the latter remains as simple as possible. We make an exception for \texttt{spark.shuffle.spill.compress}, as explained later.

\begin{table}[tb!]
\centering
\scriptsize
\caption{Average Parameter Impact}
\label{avgimpact}
\begin{tabular}{|c||c|c|c||c|}
\hline
~ & Sort-by-key & Shuffling & K-Means & Average  \\ \hline
\texttt{spark.serializer} & 26.6 \% & 9.2\% & $<$5\% & 12.6\% \\
\texttt{shuffle/storage.memoryFraction} & 13.1\% & 11.9\% &8.3\%& 11.3\% \\
\texttt{spark.reducer.maxSizeInFlight} & 5.5\% & 5.7\% & 11.5\%& 7.5\% \\
\texttt{spark.shuffle.file.buffer} & 6.3\% & 11.6\% &6.9\%& 8.2\% \\
\texttt{spark.shuffle.compress} & 137.5\% & 182\% &$<$5\%& 107.2\% \\
\texttt{spark.io.compress.codec} & $<$5\% & 18\% &6.1\%& 8.9\% \\
\texttt{spark.shuffle.consolidateFiles} & 13\% & 11\% &7.7\%& 10.5\% \\
\texttt{spark.rdd.compress} & $<$5\% & $<$5\% &5\%& $<$5\% \\
\texttt{spark.shuffle.io.preferDirectBufs} & 5.6\% & 9.9\% &$<$5\%& 5.9\% \\
\texttt{spark.shuffle.spill.compress} & $<$5\% & 6.1\% &$<$5\%& $<$5\% \\
\hline
\end{tabular}
\end{table}

%

\section{The Proposed Tuning Methodology}
\label{sec:methodology}

\begin{figure}[tb!]
\centering
\includegraphics[width=0.7\columnwidth]{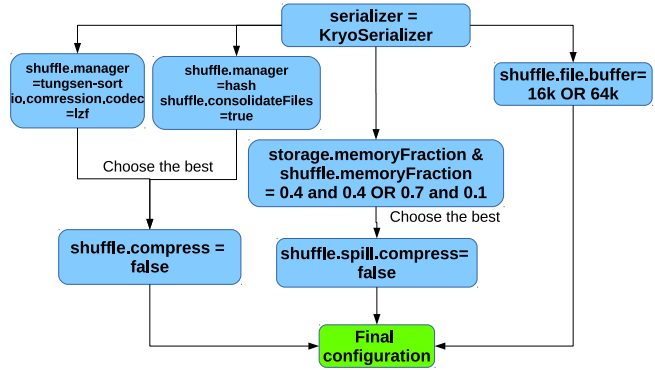}
\caption{Our proposed spark parameter tuning methodology}
\label{fig:diagram}
\end{figure}

Based on (i) the results of the previous section, (ii) the expert knowledge as summarized in Sec. \ref{sec:overview} and (iii) our overall experience from running hundreds of experiments (not all are shown in this work), we derive an easily applicable tuning methodology. This methodology is presented in Fig. \ref{fig:diagram} in the form of a block diagram. In the figure, each node represents a \emph{test run} with one or two different configurations. Test runs that are higher in the figure are expected to have a bigger impact on performance and, as a result, a higher priority. As such, runs start from the top and, if an individual configuration improves the performance, the configuration is kept and passed to its children replacing the default value for all the test runs on the same path branch. If an individual configuration does not improve the performance, then the configuration is not added and the default is kept. In other words, each parameter configuration is propagated downstream up to the final configuration as long as it  yields performance improvements. Contrary to the previous experiments, the methodology test runs investigate the combined effect of tuning.

Overall, as shown in the figure, at most ten configurations need to be evaluated referring to nine of the parameters in Sec. \ref{sec:params}. Note that, even if each parameter took only two values, exhaustively checking all combinations would result in $2^9=512$ runs.
Finally, the methodology can be employed in a less restrictive manner, where a configuration is chosen not only if it improves the performance, but if the improvement exceeds a threshold, e.g., 5\% or 10\%.

The rationale of the diagram blocks, itemized by the main parameter they target, is as follows:
\begin{itemize}
\item \texttt{spark.serializer:}  This parameter had the highest impact in our series of experiments, and using the \textit{KryoSerializer} was the default baseline for all the other parameters. We keep the same rationale in our methodology, so we perform this test  first.

\item \texttt{spark.shuffle.manager:} As shown in the results of the previous section, the shuffle manager has a high impact on performance, so it should be included in the methodology. Since, based on documentation, \textit{tungsten-sort} works better with the \textit{lzf} compression codec, we combine the test of these two settings. Also, the test run for the other option of this parameter, the \textit{hash} shuffling manager, is conducted in combination with the implementation of consolidating files during a shuffle, to avoid problems from the creation of too many intermediate files.

\item \texttt{spark.shuffle.compress:} In our parameters, disabling it led to serious performance degradation (by default it is enabled). This means that it has an impact. Interestingly, the best results presented by Spark's developers for the \emph{terasort} benchmark are produced when this is disabled, which further supports our choice to include it in our methodology.

\item \texttt{storage/shuffle.memoryFraction:}  The memory fraction allocated is inherently important in Spark, due to its main memory-oriented execution model. However, this parameter is also tightly connected to the hardware characteristics of the cluster infrastructure.

\item \texttt{spark.shuffle.spill.compress:} While this parameter appears not to have any significant impact on the performance in the experiments that we conducted, it is closely linked to the shuffling memory fraction. Since the latter is taken into consideration, we also include this one.

\item \texttt{spark.shuffle.file.buffer:} In our experiments, the impact of this parameters is rather small. However, it is included for completeness. A shorter version of our methodology with two required runs less, would omit it.

\end{itemize}

We test the effectiveness of our methodology using three case studies. More specifically, the methodology is applied to the \textit{sort-by-key}, \textit{k-means} and \textit{aggregate-by-key} benchmark applications, respectively. For the former, we keep the same input as previously. For \emph{k-means}, we use a different data input, which leads to radically different performance than previously, i.e., this instance of \textit{k-means} cannot be deemed as used for constructing our methodology.  \textit{Aggregate-by-key} is a new application.

{\bf Sort-by-key.} For \textit{Sort-by-key} over 1 billion 100-byte records, the default performance is 218 secs. We set a performance improvement threshold at 10\% of this value, i.e., 21.8 secs. The final configuration: advocated by our methodology is \texttt{spark.serializer=}  \emph{KryoSerializer} and
\texttt{shuffle.manager=} \emph{hash} and \texttt{shuffle.consolidateFiles=} \emph{true} and \texttt{shuffle/storage.memoryFraction} = 0.4/0.4: 120 secs.
Overall, the running time in this case study, decreased from 218 secs down to 120 secs (44\% performance improvement).

{\bf K-Means.} Next, we apply the methodology on \textit{K-Means} for 10 centers, 10 iterations, 100 million points and 500 columns. The runtime with the default configuration is 654 secs. The final configuration does not include the \emph{KryoSerializer} and is as follows: \texttt{shuffle/storage.memoryFraction} = 0.1/0.7 and \texttt{shuffle.\\spill.compress}=false. Overall the running time dropped by more than 91\%, from 654 to 54 secs. An interesting note is that in the profiling experiments in Section \ref{sec:exps}, \emph{k-means} did not exhibit such high speedups. This is due to the fact that the performance of \emph{k-means} is sensitive to its input dataset.

{\bf Aggregate-by-key.}
The running time for the default  configuration is 77.5 secs. As input, we use 2 billion key-value pairs with length of 10 and 90, respectively. In this case study, we set the threshold to a lower values, at 5\%.
%
%
%
For \textit{aggregate-by-key}, the overall performance improvement is about 21\%, and the configuration is
\texttt{shuffle.manager=} \emph{hash} and \texttt{shuffle.consolidateFiles=} \emph{true} and \texttt{shuffle/storage.memoryFraction} = 0.1/0.7.

\section{Conclusions}
\label{sec:concl}

This work deals with configuring Spark applications in an efficient manner. We focus on 12 key application instance-specific configurable parameters and assess their impact using real runs on a petaflop supecomputer. Based on the results and the knowledge about the role of these parameters, we derive a trial-and-error methodology, which requires a very small number of experimental runs. We evaluate the effectiveness of our methodology using three case studies, and the results show that we can achieve up to more than a 10-fold speedup. Although our results are significant, further research is required to investigate additional infrastructures, benchmark applications, parameters and combinations.


\bibliographystyle{plain}
\bibliography{sparkconf}  
%
\end{document}